\documentclass[12pt]{article}

\usepackage{amsmath,amssymb,graphicx} 
\usepackage{epsf}
\usepackage{pstricks}
\usepackage{cite}

\newcommand{\beq}{\begin{eqnarray}}
\newcommand{\eeq}{\end{eqnarray}}

\newcommand{\centeron}[2]{{\setbox0=\hbox{#1}\setbox1=\hbox{#2}\ifdim
\wd1>\wd0\kern.5\wd1\kern-.5\wd0\fi
\copy0

\kern-.5\wd0\kern-.5\wd1\copy1\ifdim\wd0>\wd1
                                       \kern.5\wd0\kern-.5\wd1\fi}}
\newcommand{\ltap}{\>\centeron{\raise.35ex\hbox{$<$}}
                               {\lower.65ex\hbox{$\sim$}}\>}
\newcommand{\gtap}{\>\centeron{\raise.35ex\hbox{$>$}}
                               {\lower.65ex\hbox{$\sim$}}\>}

\newcommand\ZZ{\hbox{\zfont Z\kern-.4emZ}}
\font\zfont = cmss10 

\begin{document}
\begin{titlepage}
\begin{flushright}
\end{flushright}

\vskip.5cm
\begin{center}
{\huge \bf Nearby resonances beyond \\
\vskip0.2cm
the Breit-Wigner approximation}

\vskip.1cm
\end{center}
\vskip0.2cm

\begin{center}
{\bf
{Giacomo Cacciapaglia}$^{a}$, {Aldo Deandrea}$^{a}$,
{\rm and}
{Stefania De Curtis}$^{b}$}
\end{center}
\vskip 8pt

\begin{center}
$^{a}$ {\it Universit\'e de Lyon, F-69622 Lyon, France; Universit\'e Lyon 1, Villeurbanne;\\
CNRS/IN2P3, UMR5822, Institut de Physique Nucl\'eaire de Lyon\\
F-69622 Villeurbanne Cedex, France } \\
\vspace*{0.1cm}
$^{b}$ {\it INFN, 50019 Sesto Fiorentino, Firenze, Italy} \\
\vspace*{0.1cm}

\end{center}

\vglue 0.3truecm

\begin{abstract}
\vskip 3pt \noindent We consider a description of propagators for
particle resonances which takes into account the quantum mechanical
interference due to the width of two or more nearby states that have
common decay channels, by incorporating the effects arising from the
imaginary parts of the one-loop self-energies. Depending on the
couplings to the common decay channels, the interference effect, not
taken into account in the usual Breit-Wigner approximation, can
significantly modify the cross section or make the more long-lived
resonance narrower. We give few examples of New Physics models for
which the effect is sizable, namely a generic two and multiple Higgs model and  neutral vector resonances in Higgsless
models. Based on these results we suggest the implementation of a
proper treatment of nearby resonances into Monte Carlo generators.
\end{abstract}

\end{titlepage}

\newpage


\section{Introduction}
\label{sec:intro}
\setcounter{equation}{0}
\setcounter{footnote}{0}

The Breit-Wigner (BW) approach to resonances in particle physics allows
to take into account the finite width of a meta-stable particle.
Various forms of this approach exist, which allow to describe
different  situations, from the narrow width approximation, up to
broad and energy dependent widths.

In the following we consider a generalisation of the Breit-Wigner
description \cite{Pilaftsis:1997dr} which makes use of a matrix propagator including
non-diagonal width terms in order to describe physical examples in which these effects are relevant.
Indeed for more than one meta-stable state
coupled to the same particles, loop effects will generate mixings
for the masses as well as mixed contributions for the widths
(imaginary parts). In general a diagonalisation procedure for the
masses (mass eigenstates) will leave non-diagonal terms for the
widths.

Usually non-diagonal width terms are discarded. This is  a good
approximation in most cases and different unstable particles are
described each by an independent Breit-Wigner profile. However when
two or more resonances are close-by and have common decay channels
(a precise definition of nearby resonances will be given in the
following) such a description is not accurate anymore. Indeed, already
from a quantum mechanical point of view, one cannot treat these states
as independent. The usual Breit-Wigner approximation amounts to sum
the modulus square of the various amplitudes neglecting the
interference terms. When there are common decay channels and the
widths of the unstable particles are of the same order of the mass splitting, the interference terms
may be non-negligible. Generalisations of the Breit-Wigner approach
were discussed in the literature in the past (see \cite{Pilaftsis:1997dr} for a general
discussion of unstable-particle mixing, gauge invariance issues, unitarity and renormalisation
for strongly-mixed systems), mainly focussing to the applications in the Higgs sector of the
supersymmetric standard model and CP violation
\cite{Ellis:2004fs,Frank:2006yh,Hahn:2007it,Dreiner:2007ay,Kittel:2008be}.
In the following we shall give a general formalism for scalar and vector resonances and
discuss other relevant physical examples. In particular we will
consider models of physics Beyond the Standard Model (BSM) in which new
resonances play a crucial role, such as a generic two-Higgs model or
the case of an arbitrary number of Higgs particles. Interesting is
also the case of Higgsless models for which the lowest lying Kaluza-Klein (KK)
excitations of the photon and the $Z$ are nearly degenerate and have
common decay channels into fermions and gauge bosons. We will show
that in all these cases, the interference effects can play an
important role. Based on these results we suggest that a proper
treatment  should be carefully implemented into Monte Carlo
generators as physical results may be dramatically different from a
naive use of the Breit-Wigner approximation.

\section{Scalar fields}
\label{sec:scalar}
\setcounter{equation}{0}

We first discuss the case of scalar fields which gives a simpler
overview of the problem  without the extra complications of the
gauge and Lorentz structure present in other cases. The main
observation needed to take the mixing effect into account, is to
remember the link between the action and the propagator in quantum
field theory and write the correction to the propagator as a
modification of the kinetic operator, which, in the general case, is
a matrix.

Let's start with the  action for a real scalar field given by (in
position and momentum space):
\beq \mathcal{L}_{\rm scalar} = \int d^4 x\; \frac{1}{2} \left[
(\partial_\mu \phi)^2 - m_0^2 \phi^2 \right] =  \int \frac{d^4 p}{(2
\pi)^4}\; \frac{1}{2} \phi (-p)  K_{s,0} (p) \phi (p)\,, \eeq where,
for later convenience, we have defined a kinetic function
\beq K_{s,0} (p) = p^2 - m_0^2\,. \eeq The propagator of the scalar
field is defined as the inverse of the kinetic operator:
\beq i \Delta_{s,0} = i K_{s,0}^{-1} = \frac{i}{p^2 - m_0^2}\,. \eeq
In the presence of interactions, the kinetic term will receive
contributions by loops: if we call $i \Pi (p^2)$ the value of the
1-PI corrections to the propagator, the corrected kinetic term is
\beq K_s = p^2 - m_0^2 + \Pi (p^2)\,. \eeq
Therefore the new
propagator is
\beq i \Delta_s = i K_s^{-1} = \frac{i}{p^2-m_0^2+\Pi (p^2)} = i
\Delta_{s,0} \sum_{n=0}^\infty (-1)^n \left(\Pi  \Delta_{s,0}
\right)^n\,, \eeq
which corresponds to the resummation of the 1-PI
insertions on the bare propagator. The pole of the resummed
propagator defines, as usual, the renormalised mass. If the particle
is unstable, $\Pi(p^2)$ is complex. The real part is used to
renormalise the mass and the imaginary part defines the width of the
particle. The propagator has a pole in $ m^2 - i m \Gamma$, with $m$
the renormalised mass and $\Gamma$ the width. For our purposes, we
leave out  the imaginary part
\beq
\Im \Pi (p^2)= \Sigma (p^2)\,.
\eeq
In the narrow width approximation, $\Sigma (m^2) =  m \Gamma$.

This formalism can be generalised to a system involving
multi-fields, which do couple to the same intermediate  particles:
the loops will generate mixings in the masses, but also out-of-diagonal imaginary parts.
In general the real and imaginary parts
will not be diagonalisable at the same time: we are interested in
the phenomenological consequences of this scenario, when the
out-of-diagonal imaginary part is of the same order as the mass
splitting. The kinetic function is now  written in matrix form:
\beq
(K_s)_{lk} = (p^2 - m_l^2) \delta_{lk} +  i \Sigma_{lk} (p^2)\,.
\eeq
(We are considering the imaginary part only, the
real one is used to renormalise the masses.) The propagator of the
fields can be defined as the inverse of this matrix:
\beq i (\Delta_s)_{lk}  = i \left( K_s^{-1} \right)_{lk} \,.
\label{delta}\eeq

To give an explicit example, let us focus on the two-particle case:
\beq \label{eq:twoscalar}
i \Delta_s = \frac{i}{D_s} \left( \begin{array}{cc}
p^2 - m_2^2 + i \Sigma_{22} & -i \Sigma_{12} \\
-i \Sigma_{21} & p^2 - m_1^2 + i \Sigma_{11}
\end{array} \right)\,,
\eeq where \beq D_s = (p^2 - m_1^2 + i \Sigma_{11})(p^2 - m_2^2 + i
\Sigma_{22} ) + \Sigma_{12} \Sigma_{21}\,. \eeq For vanishing
$\Sigma_{12}$ and $\Sigma_{21}$, the propagator is diagonal and  it
reduces to two independent Breit-Wigner propagators with $m_i
\Gamma_i = \Sigma_{ii} (m_i^2)$.

However, the narrow width approximation is not valid if the off-diagonal terms are sizable compared with the mass splitting.
Defining $2 M^2 = m_2^2+m_1^2$
and $2 \delta = m_2^2 - m_1^2$, the poles of the propagator (zeros
of $D_s$), which define the physical masses and widths of the two resonances,
are:
\beq \tilde{m}_{\pm}^2 = M^2 - i \frac{\Sigma_{11} + \Sigma_{22}}{2}
\pm \frac{i}{2} \sqrt{(\Sigma_{22} - \Sigma_{11} + 2 i \delta)^2 + 4
\Sigma_{12} \Sigma_{21}}\,. \eeq Note that the value of the masses
is modified by the presence of the off-diagonal terms due to the
imaginary part of the square root, at the same time the widths are
affected. More importantly, the off-diagonal terms in the propagator
will generate non-negligible interference, which can be in turn
constructive or destructive. This effect will be illustrated with
some numerical examples.

Finally, let us write some general formulae for $\Sigma(p^2)$. If we
assume that the scalar particles $i$ and $j$  couple to a pair of
particles $\alpha$ with couplings $\lambda_{\alpha}^i$ and
$\lambda_{\alpha}^j$ respectively, the matrix $\Sigma_{ij}$ can be
written in general as
\beq [\Sigma(p^2)]_{ij} = \sum_{\alpha}
\lambda_{\alpha}^i\lambda_{\alpha}^j f_\alpha (p^2)\,.
\label{sigma}\eeq We will consider couplings with fermions $f$,
scalars $s$ and vectors $V$:
\beq \phi_i \left( \bar{f} (\lambda_{fL}^i P_L + \lambda_{fR}^i P_R)
f' + \lambda_s^i s^\dagger s' + \lambda_V^i V^\dagger_{1\mu} V_2^\mu
\right)\,, \eeq where $P_{L,R}= (1\mp \gamma_5)/2$ are the chirality
projectors. Therefore, expanding for small masses of the particles
in the loop (full results are given in appendix A), we get:
\beq
\left[\Sigma^{(s)}_f (p^2)\right]_{ij} &=& \frac{\lambda^i_{fL} \lambda^j_{fR} + \lambda^i_{fR} \lambda^j_{fL}}{16 \pi} p^2 + \dots \\
\left[\Sigma^{(s)}_s (p^2)\right]_{ij} &=& \frac{\lambda_s^i \lambda_s^j}{16 \pi} + \dots \\
\left[\Sigma^{(s)}_V (p^2)\right]_{ij} &=&  \frac{\lambda_V^i
\lambda_V^j}{64 \pi} \frac{p^4}{m_{V_1}^2 m_{V_2}^2} + \dots
\label{eq:scalq}\eeq

Note that the momentum dependence in the previous formula
could give rise to violation of high-energy unitarity at
energies above the resonance as well as distortion of the line-shape
for broad resonances. These issues are discussed in detail in
\cite{Pilaftsis:1997dr,Papavassiliou:1997pb}. Actually we do not
face these problems here, since we are interested in the effects of
the absorptive self-energies near the pole of nearby resonances
whose widths are of the same order of magnitude of the mass
splitting.

\subsection{A numerical example: near-degenerate Higgses}

As a numerical example, we will study two heavy Higgses where both
the scalars develop a vacuum expectation value (VEV) and therefore couple to the $W$ and $Z$
gauge bosons. This situation is common in supersymmetric models
where two Higgses are required by writing supersymmetric Yukawa
interactions for up and down type fermions, and generic two Higgs
models. The interference between near degenerate Higgses has been
studied in \cite{Ellis:2004fs,Frank:2006yh,Hahn:2007it} focusing in
CP violation effects.

The couplings of the two CP-even Higgses to gauge bosons can be written as
\beq
\lambda_{WWH1} = g\, m_W \cos \alpha\,, &\qquad& \lambda_{WWH2} = g\, m_W \sin \alpha\,, \nonumber\\
\lambda_{ZZH1} = \frac{g\, m_Z}{\cos \theta_W} \cos \alpha\,,
&\qquad& \lambda_{ZZH2} = \frac{g\, m_Z}{\cos \theta_W} \sin
\alpha\,; \label{couplings}\eeq where $\alpha$ is a mixing angle
taking into account the mixing between the two mass eigenstates and
the difference between the two VEVs.

\begin{figure}[tb]
\begin{center}
\includegraphics[width=14cm]{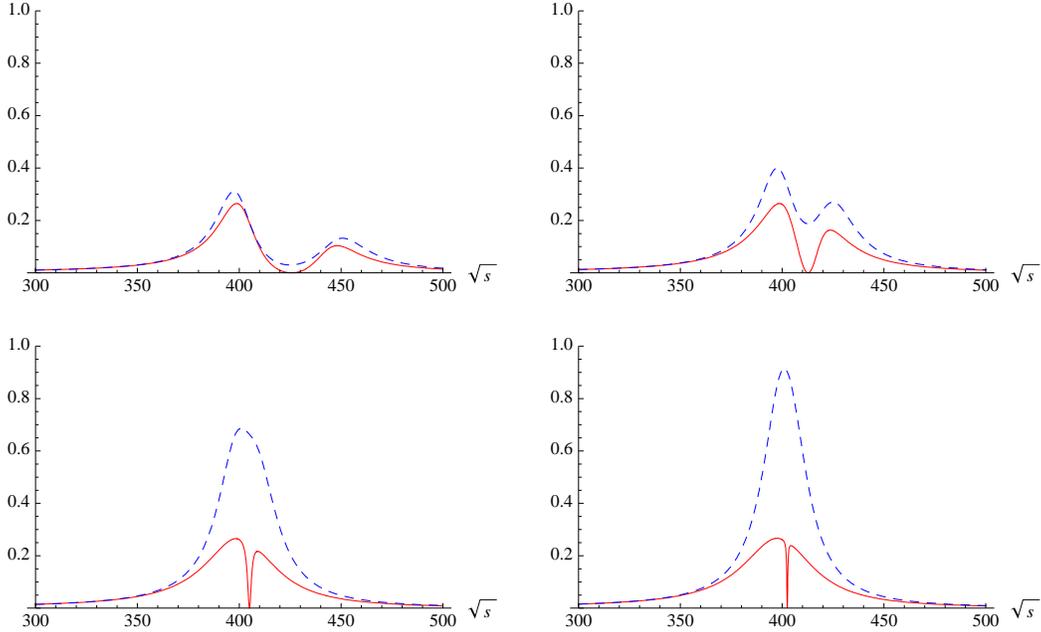}
\end{center}
\caption{\footnotesize Plots of the production cross section (in
arbitrary units) of two nearby Higgses decaying into gauge boson
pairs for the naive Breit-Wigner (blue-dashed) and exact mixing
(red-solid). The mass of the first resonance is fixed to $400$ GeV,
the splitting respectively 50, 25, 10 and 5 GeV and $\alpha=\pi/4$.}
\label{fig:gnomettiHiggs}
\end{figure}

Here we are interested in a generic production cross section of the
two nearby Higgses on the resonances, with decay of the Higgses into
gauge bosons (either $WW$ or $ZZ$). The amplitude of this process is
proportional to the resonant propagator weighted by the couplings
given in eq.(\ref{couplings}). In the case we are considering, the
common decay channels can give off-diagonal terms in
eq.(\ref{eq:twoscalar}) which are sizable compared with the mass
splitting. Therefore we need to include their effects and  a generic
cross section will be proportional to (here we assume that the
coupling to the initial particles are the same):
\beq
\left|(\Delta_s^{11} + \Delta_s^{21}) \cos \alpha + (\Delta_s^{22}+ \Delta_s^{12} ) \sin \alpha \right|^2\,.
\eeq
In Fig.~\ref{fig:gnomettiHiggs}, we plot this quantity in arbitrary units and compare it with the
Breit-Wigner approximation:
 we fix $m_{H1} = 400$ GeV, and vary the splitting from
50 to 5 GeV. For simplicity, in the following we will assume $\alpha
= \pi/4$, so that the two scalars have the same couplings (but this
assumption is not crucial for our conclusions). The exact treatment
of the resonances unveils a destructive interference (which is
neglected in the naive Breit-Wigner case) that  can drastically
reduce the cross section. Also, the interference between the two
resonances splits the mass poles \cite{Frank:2006yh,Hahn:2007it}.

\begin{figure}[tb]
\begin{center}
\includegraphics[width=14cm]{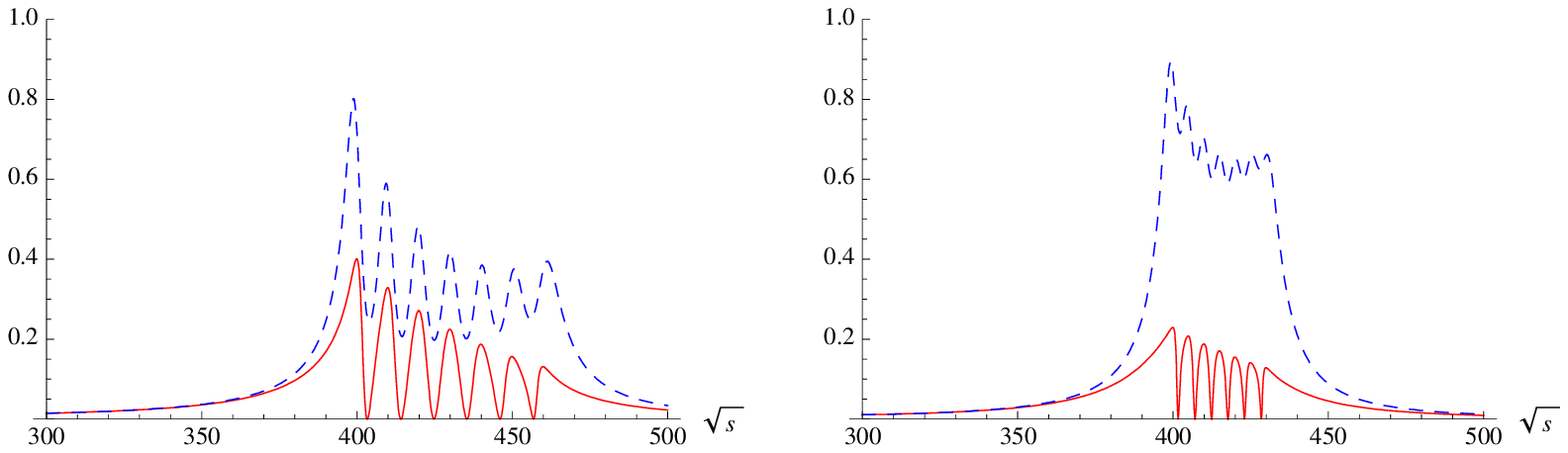}
\end{center}
\caption{\footnotesize Plots of the production cross section (in
arbitrary units) for seven nearby Higgses equally coupled to SM gauge
bosons: the naive Breit-Wigner (blue-dashed) bump reduces to a row of seven dwarfs when the exact mixing
(red-solid) is taken into account. The mass of the first resonance is fixed to $400$ GeV,
the splitting between the six Higgses respectively 10 and 5 GeV.}
\label{fig:gnometti}
\end{figure}

This effect can be even more important for scenarios with a large
number of scalars as predicted in some string models. Our analysis
can be easily extended to an arbitrary number of Higgses. Let's take
for example the couplings to the gauge bosons  to be given by
$g_{\rm SM}/\sqrt{N}$, where $g_{SM}$ is the SM coupling of the
gauge bosons and $N$ is the number of Higgses. In
Fig.~\ref{fig:gnometti} we plot the cross section for seven nearby
Higgses, with the first one at 400 GeV and the others at a distance
of 5 and 10 GeV, the width of each being 6.2 GeV. From the plot it
is clear that the destructive interference reduces the giant
resonance (which is not distinguishable from a single Higgs, once
the experimental smearing is taken  into account) to a bunch of
{\it gnometti} (dwarfs), which will be very hard to detect. The cross
section is in fact reduced by a significant factor with respect to the
naive expectation, and the smearing will wash out the peak
structure. This situation is different from the continuum Higgs
spectrum of \cite{Espinosa:1998xj}, where the Higgs peaks are
smeared by the experimental uncertainties only. Therefore, in this
case, together with the appearance of a ``continuum'', the cross
section is suppressed by the interference. It is intriguing to
compare this analysis with Un-Higgs
models~\cite{Stancato:2008mp,Falkowski:2008yr}, where the Higgs in
indeed a continuum: such behaviour may arise from the superposition
of Kaluza-Klein resonances in extra dimensional realisations or
deconstructed models~\cite{Falkowski:2008yr}.

\section{Vector fields}
\label{sec:vector}
\setcounter{equation}{0}

For a vector field the technique is similar to the scalar case,
however a major issue is  defining the propagator of a metastable
particle in a gauge invariant way (see for example
\cite{Nowakowski:1993iu}). Nevertheless, at the pole, only the values of
the pole mass and width (which are gauge-invariant) are relevant,
and extra terms that must be added to preserve gauge invariance are
numerically negligible. The kinetic function for a vector $V$, in
generic $\xi$-gauge, is
\beq
K_{\mu\nu,0} &=& g_{\mu \nu} (p^2 - M_0^2) - p_\mu p_\nu \left(\frac{1}{\xi} -1 \right) \nonumber\\
 & = & \left( g_{\mu \nu} - \frac{p_\mu p_\nu}{p^2} \right)  (p^2 - M_0^2) + \frac{p_\mu p_\nu}{p^2} \left(
\frac{p^2}{\xi} - M_0^2 \right)\,. \eeq
The propagator, is then
defined as the inverse of the kinetic term:
\beq i \Delta_{\mu\nu,0} &=& -i  \left( g_{\mu \nu} - \frac{ p_\mu
p_\nu}{p^2} \right)
(p^2 - M_0^2)^{-1} - i \frac{p_\mu p_\nu}{p^2} \left( \frac{p^2}{\xi} - M_0^2 \right)^{-1} \nonumber\\
 & = &  \left( g_{\mu \nu} - \frac{p_\mu p_\nu}{p^2} \right)  \frac{-i}{p^2 - M_0^2} + \frac{p_\mu p_\nu}{p^2}
\frac{ -i \xi}{p^2 -\xi M_0^2}\,. \eeq
The first term of the
propagator has a gauge-independent pole at the $V$ mass, while the
other part has a gauge-dependent pole, which will cancel the pole
given by the Goldstone boson (whose mass is indeed $\xi M_0^2$).
Therefore, at the pole we can neglect the contribution of the second
term, and the propagator simplifies to
\beq i \Delta_{\mu\nu,0} \simeq \left( g_{\mu \nu} - \frac{p_\mu
p_\nu}{p^2} \right)  \frac{-i}{p^2 - M_0^2} \,; \eeq
which is equal to a Lorentz tensor times a scalar propagator.

Factorising out the Lorentz structure, loop corrections can be
parameterised as
\beq
\Pi_{\mu \nu} = \Pi_T\, g_{\mu \nu} + \Pi_L\, p_\mu p_\nu\,,
\eeq
so that
\beq K_{\mu\nu,0} + \Pi_{\mu \nu} =  \left( g_{\mu \nu} -
\frac{p_\mu p_\nu}{p^2} \right)  (p^2 - M_0^2 + \Pi_T) + \frac{p_\mu
p_\nu}{p^2} \left( \frac{p^2}{\xi} - M_0^2  + \Pi_T + p^2
\Pi_L\right)\,. \eeq The corrected propagator is therefore:
\beq i \Delta_{\mu \nu} =  \left( g_{\mu \nu} - \frac{p_\mu
p_\nu}{p^2} \right)  \frac{-i}{p^2 - M_0^2 + \Pi_T} + \frac{p_\mu
p_\nu}{p^2} \frac{ -i \xi}{p^2 -\xi (M_0^2 - \Pi_T - p^2 \Pi_L)}\,.
\eeq
The first term defines the pole mass and width (gauge
independent), while the second term contains a gauge-dependent pole,
and it is negligible at the physical pole. Here we will be
interested only in the imaginary part  $\Im \Pi_T = \Sigma$, which
defines the decay width of the vectors. Neglecting the second part
of the propagator, we find:
\beq
i \Delta_{\mu \nu} =  \left( g_{\mu \nu} - \frac{p_\mu p_\nu}{p^2} \right) (-i)
\left( p^2 - M_V^2 + i \Sigma \right)^{-1} =  \left( g_{\mu \nu} - \frac{p_\mu p_\nu}{p^2} \right) (-i) \Delta_s (M_V)\,.
\eeq
For a generic number of vectors, the same discussion as in the
scalar case applies, and $\Delta_s $ has a matrix form, like in
(\ref{delta}).

We conclude with some general formulae for $\Sigma$ which can be
expressed as in eq. (\ref{sigma}). We will consider couplings with
fermions $f$, scalars $s$ and vectors $V$:
\beq V_i^\mu \left( \bar{f} \gamma_\mu (\lambda^i_L P_L +
\lambda^i_R P_R) f' +  \lambda_s^i (q_1 - q_2)^\mu s_1^\dagger s_2 +
\lambda_V^i V_1^\nu V_2^\rho G_{\mu \nu \rho} \right)\,, \eeq where
$G_{\mu \nu \rho} = g_{\mu \nu} (p + q_1)_\rho + g_{\nu \rho} (q_2 -
q_1)_\mu - g_{\rho \mu} (q_2 + p)_\nu$, with $p = q_1 + q_2$.
Expanding for small masses of the particles in the loop (full
results are given in Appendix A) we obtain:
\beq
\left[\Sigma^{(V)}_f (p^2)\right]_{ij} &=& \frac{\lambda^i_L \lambda^j_R + \lambda_R^i \lambda_L^j}{24 \pi} p^2 + \dots \\
\left[\Sigma^{(V)}_s (p^2)\right]_{ij} &=& \frac{\lambda_s^i \lambda_s^j}{48 \pi} p^2 +\dots \\
\left[\Sigma^{(V)}_V (p^2)\right]_{ij} &=& \frac{\lambda_V^i
\lambda_V^j}{192 \pi} \frac{p^6}{m_{V_1}^2 m_{V_2}^2}+ \dots \eeq
In the following we apply this formalism to a simple case.

\subsection{Numerical example: $Z'$ and $A'$ in Higgsless models}

In Higgsless models \cite{Csaki:2003zu,Cacciapaglia:2004rb}, the first two neutral resonances are
nearly degenerate, and they correspond to the first KK excitation
of the $Z$ and of the photon. Here we will explicitly refer to the warped extra dimensional model in \cite{Cacciapaglia:2004rb}: the masses can be approximated by
\beq m_{Z'}^2 \simeq m_{KK}^2 + 4 m_Z^2\,, \qquad m_{A'}^2 \simeq
m_{KK}^2\,, \eeq so that the mass difference is very small:
\beq
m_{Z'} - m_{A'} \simeq 2 \frac{m_Z^2}{m_{KK}} \sim 16\, {\rm GeV} \cdot \left( \frac{1 {\rm TeV}}{m_{KK}} \right)^2\,.
\eeq
In terms of the parameters of the warped geometry ($R$ is the curvature, $R'$ the position of the Infra-Red brane in covariant coordinates):
\beq
m_{KK} \sim \frac{2.4}{R'}\,, \qquad m_W = \frac{1}{R' \log \frac{R'}{R}}\,;
\eeq
therefore, given the value of the curvature $R$, the KK mass ($R'$) is determined by the $W$ mass.

The coupling to the light gauge bosons can be estimated by use of
the Higgsless sum rules, that ensure the cancellation of the terms
growing like the energy square in the elastic scattering amplitude of the
longitudinal $W$ and $Z$, thus delaying the violation of perturbative
unitarity at higher scales than in the Standard Model without Higgs. One of these sum rules is \cite{Csaki:2003dt,Cacciapaglia:2006mz}:
\beq \label{eq:sumrule}
g_{WWWW} = \frac{3}{4} \left( g_{WWZ}^2 \frac{m_Z^2}{m_W^2} +
\sum_{k} g_{WWk}^2 \frac{m_k^2}{m_W^2} \right) + \frac{g^2}{4}
(1-\zeta)\,, \eeq
where $g_{WWWW} = g^2$, $g_{WWZ} = g \cos
\theta_W$ and we have included a partial contribution for the Higgs
($\zeta = 1$ corresponds to the Higgsless limit, $\zeta \to 0$ to
the SM Higgs). Neglecting the contribution of the heavier states, from eq.(\ref{eq:sumrule}) we can estimate the coupling of the first tier:
\beq g_{WW1} \simeq \sqrt{\frac{\zeta}{3}} g \frac{m_{W}}{m_{KK}}\,.
\eeq This is actually the coupling of the first KK mode of the
neutral component in the SU(2) multiplet: therefore, in order to
obtain the couplings of the mass eigenstates, one needs to impose a
rotation by an angle $\theta_1$, which describes the mixing between
the SU(2) and the U(1) vectorial component:
\beq g_{WWZ'} \simeq g_{WW1}  \cos \theta_1\,, \qquad g_{WWA'}
\simeq g_{WW1}  \sin \theta_1\,. \eeq Numerically, it turns out that
the $Z'$ is 45\% in the SU(2) and 55\% in U(1), therefore $\cos
\theta_1 \sim \sqrt{45\%}\sim 0.67$ and $\sin \theta_1 \sim
\sqrt{55\%} \sim 0.74$.

The amplitude for the decay $Z', A' \to WW$ is given by
($\alpha_W=g^2/4\pi$):
\beq
\Sigma (p^2) \simeq \left( \begin{array}{cc}
\cos^2 \theta_1 & \cos \theta_1\, \sin \theta_1 \\
 \cos \theta_1\, \sin \theta_1 & \sin^2 \theta_1 \end{array} \right) \frac{\zeta \alpha_W}{144} \frac{p^6}{m_W^2 m_{KK}^2}\,;
\eeq
and the width can be estimated by:
\beq
\Gamma = \frac{\Sigma (m_{KK}^2)}{m_{KK}} \sim \left( \begin{array}{cc}
17 & 19 \\
19 & 21 \end{array} \right){\rm GeV} \cdot \left( \frac{m_{KK}}{1
{\rm TeV}} \right)^3 \zeta\,. \eeq
The off-diagonal entries are large, and they are of the same order of the mass splitting between the two neutral gauge bosons.

The determination of the couplings to fermions, which are relevant
for the Drell-Yan production, is more involved. In fact, the typical
scenario is that the right-handed components of the light quarks and
leptons are localised on the Ultra-Violet brane, while the left-handed
components are spread in the bulk~\cite{Cacciapaglia:2004rb}. For
the left-handed couplings, we have some freedom: however, electroweak precision measurements
prefer small couplings therefore we will neglect them. The couplings
of the right-handed components depend uniquely on the suppression of
the wave functions on the UV brane. Therefore, we can approximate:
\beq g_{f \bar{f} 1} \simeq Q_f g\, \tan \theta_W\;
\frac{1}{\sqrt{\log \frac{R'}{R}}} \sim Q_f g\, \tan \theta_W\; 2.4
\frac{m_W}{m_{KK}} \sim Q_f \cdot 0.066 \cdot \left( \frac{1 {\rm
TeV}}{m_{KK}} \right)\,. \eeq Numerically we find that:
\beq g_{f \bar{f} Z'} = g_{f \bar{f} 1}\,, \qquad g_{f \bar{f} A'} =
g_{f \bar{f} 1} \eta\,; \eeq were the ratio $\eta \sim -0.32$ is to
a good approximation independent on the value of the KK mass. In
Table 1, three numerical examples are given, corresponding to
different values of the curvature: the masses and couplings are
calculated exactly, following \cite{Csaki:2003zu}, and confirm our
estimates.

\begin{table}[t]
\centering
\begin{tabular}{|c|ccc|cccc|}
\hline
$R$  & $m_{W'}$ & $m_{A'}$ & $m_{Z'}$ &  $g_{WWA'}$ & $g_{WWZ'}$ & $g_{f \bar{f} A'}$ & $g_{f \bar{f} Z'}$  \\
\hline
$10^{-15}$ & 1032 & 1028.5 & 1043.79 & 0.022 & 0.019 & -0.021 & 0.065 \\
$10^{-10}$ & 805.3 & 801.0 & 820.9 & 0.028 & 0.024 & -0.027 & 0.085 \\
$10^{-7}$ & 634.6 & 629.0 & 655.0 & 0.037 & 0.029 & -0.035 & 0.11 \\
\hline
\end{tabular}
\caption{Three Higgsless points for the model proposed in
\cite{Csaki:2003zu}, masses in GeV.} \label{tab:higgsless}
\end{table}

We are interested in three processes: Drell-Yan production and decay
into gauge bosons $W^+ W^-$ (DY), Drell-Yan production and decay
into a pair of leptons (Leptonic) and vector boson fusion production
followed by decay into gauge bosons (VBF).
As in the scalar case, the amplitudes of these processes on resonance are proportional to the propagators weighted by the couplings with the incoming and outcoming particles.
The cross section $\sigma
(q \bar{q} \to A',Z' \to W^+ W^-)$  is proportional to:
\beq \sigma_{\rm DY} \simeq \left|\cos \theta_1\,\Delta_s^{Z'Z'} +
\sin \theta_1 \Delta_s^{Z'A'} - |\eta| ( \sin \theta_1
\Delta_s^{A'A'} + \cos \theta_1\, \Delta_s^{A'Z'} )  \right|^2\,,
\eeq which must be compared to the naive sum of two Breit-Wigner
distributions:
\beq
\sigma_{\rm DY}^{\rm naive} \simeq \left|\cos \theta_1\, \Delta_s^{Z'Z'} - |\eta| \sin \theta_1\, \Delta_s^{A'A'}\right|^2\,.
\eeq
On the other hand,  the leptonic cross section $\sigma (q \bar{q}
\to A',Z'\to l^+l^-)$ is proportional to:
\beq
\sigma_{\rm Leptonic} &\simeq& \left|\eta^2\, \Delta_s^{A'A'} + \Delta_s^{Z'Z'} - |\eta|\; ( \Delta_s^{Z'A'} + \Delta_s^{A'Z'})  \right|^2\,; \\
\sigma_{\rm Leptonic}^{\rm naive} &\simeq&  \left| \eta^2\,
\Delta_s^{A'A'} + \Delta_s^{Z'Z'}\right|^2\,. \eeq Finally, for the
VBF channel we have:
\beq
\sigma_{VBF} &\simeq& \left|\sin^2 \theta_1\; \Delta_s^{A'A'} + \cos ^2 \theta_1\; \Delta_s^{Z'Z'} + \cos \theta_1\, \sin \theta_1\; ( \Delta_s^{Z'A'} + \Delta_s^{A'Z'})  \right|^2\,; \\
\sigma_{VBF}^{\rm naive} &\simeq&  \left| \sin^2
\theta_1\;\Delta_s^{A'A'} + \cos ^2 \theta_1\;
\Delta_s^{Z'Z'}\right|^2\,. \eeq
DY and Leptonic cross sections  are easily calculable;
the VBF channel
requires a numerical study or a more involved approximate analytical
expression.

\begin{figure}[tb]
\begin{center}
\includegraphics[width=16cm]{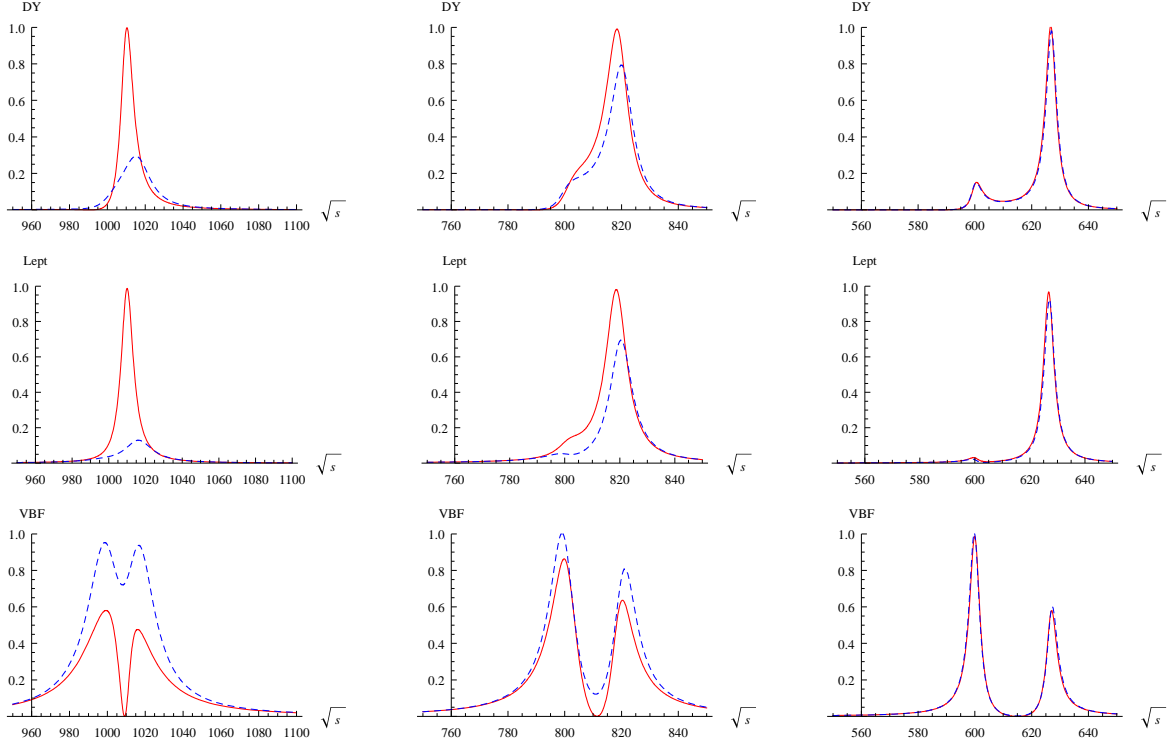}
\end{center}
\caption{\footnotesize Plots of production cross section (in
arbitrary units) of the two low-lying neutral resonances of the
Higgsless model for the naive Breit-Wigner (blue-dashed) and exact
mixing (red-solid). The rows correspond (from top to bottom) to DY,
Leptonic and VBF; the columns (from left to right) correspond to
$m_{KK} = 1000$ GeV, $800$ GeV and $600$ GeV.}
\label{fig:gnomettiHless}
\end{figure}

In  Figure \ref{fig:gnomettiHless} we plot, for illustrative
purposes, the squared matrix element of the three resonant
production channels for $A'$ and $Z'$ as function of $\sqrt{s}$ for
 three different cases: $m_{KK} = 1000$ GeV, $800$ GeV and $600$
GeV. For large masses, the effect of the interference is very
important and it can affect the value of the cross section
significantly. To make this more clear, we give hereafter  the ratio
of the area under the peaks in the figure  obtained by the exact
formula and the BW case. This roughly corresponds to the ratio of
the integrated cross sections.
\begin{center}
\begin{tabular}{cccc}
\hline
$M_{KK}=$ & 1000 GeV & 800 GeV & 600 GeV \\
\hline
DY: & 1.6 & 1.15 & 1.02 \\
Lept: & 3.15 & 1.4 & 1.05 \\
VBF: & 0.6 & 0.8 & 0.97 \\
\hline
\end{tabular} \end{center}
In the VBF channel there can be a reduction up to 50\%, while in the
other two channels the interference is constructive and the total
cross section can be enhanced by a factor of 2--3. The interference
is therefore extremely important, especially in the TeV region.
Since this represents the upper bound for   Higgsless models, the
interference effects are   crucial to determine if the whole
Higgsless parameter space can be probed at the LHC. As a consequence
the implementation of the exact matrix propagator in the Monte Carlo
generators seems mandatory for a correct analysis.

\section{Conclusions}
\label{sec:conclusions}
\setcounter{equation}{0}

We have shown that for two or more unstable particles, when there
are common decay  channels and the masses are nearby, the
interference terms may be non-negligible.  This fact is well known in
the literature and well studied since more than a decade \cite{Pilaftsis:1997dr}, here we provided
further examples in which this formalism is important.
This kind of scenario is not
uncommon in models of New Physics beyond the Standard Model,
especially in models of dynamical electroweak symmetry breaking or
in extended Higgs sectors. We reviewed the formalism for scalar and
vector fields based on a matricial form of the propagator for
multi-particles taking care of this case. This formalism can be
easily extended to fermionic resonances as discussed in \cite{Pilaftsis:1997dr} and applied to
heavy neutrino mixing in \cite{Bray:2007ru}. We gave few examples in
which the effect of the non-diagonal width is important in physical
results. In models with multi-Higgses and in Higgsless models with
near degenerate neutral vector resonances, we showed that
interference induced by the off-diagonal propagators are very
important and they can either suppress or enhance the total cross
sections on resonance depending on the relative sign of the couplings to the initial and final states.
Similar issues were discussed in the CP violating MSSM for the Higgs sector in \cite{Ellis:2005fp}
and \cite{Bernabeu:2006zs}. We expect similar effects in all the New
Physics schemes involving  resonances for which the mass splitting
is of the same order of the off-diagonal matrix elements in the
propagator due to common decay channels. For example in generic
Technicolour models in which vector and axial-vector composite states
can have the aforementioned property or also in supersymmetric
models with two nearby neutralinos.
Other examples are models in warped extra dimension, like gauge-phobic Higgs models and
Composite Higgs models.

As a conclusion, the interference effects can be crucial to study
the phenomenology of such models at the LHC, and to determine its
discovery potential. We stress that a proper treatment should be
carefully and systematically implemented into Monte Carlo generators
used to study BSM models, as physical results may be dramatically
different from a naive use of the Breit-Wigner approximation.

\section*{Acknowledgements}
We thank Veronica Sanz and Adam Martin whose discussions triggered
this project and accompanied us during the completion.
S.D.C. thanks IPNL for its hospitality while part of this work was
completed. We would also like to thank the Les Houches Center for Physics
and the Physics at TeV Colliders workshop
for hospitality during the final completion of the manuscript.
The research of G.C. and A.D. is supported in part by the
ANR project SUSYPHENO (ANR-06-JCJC-0038).

\appendix

\section{Appendix: exact amplitudes}
\label{app:ampl}
\setcounter{equation}{0}

We give in this appendix some detailed formulae which were not given or given in approximate form in the text.
Defining:
\beq \lambda (p, m_A, m_B) = \frac{(p^2 - m_B^2 + m_A^2)^2 - 4 p^2
m_A^2}{p^4}\,, \eeq the imaginary contribution to of the 1-PI
corrections of the scalar amplitudes are, at one-loop level, ($m_A$
and $m_B$ are the masses of the particles in the loop):
\beq
\Sigma_f^{(s)}(p^2)& =& \frac{\sqrt{\lambda (p, m_A, m_B)}}{16 \pi} \left[ (\lambda_L^i \lambda_R^j + \lambda_R^i \lambda_L^j )(p^2 - m_A^2 - m_B^2) + \right. \nonumber \\
& & \left. - 2 (\lambda_L^i \lambda_L^j + \lambda_R^i \lambda_R^j ) m_A m_B \right]\,, \\
\Sigma_s^{(s)}(p^2) &=&  \frac{\sqrt{\lambda (p, m_A, m_B)}}{16 \pi}  \lambda_s^i \lambda_s^j\,, \\
\Sigma_V^{(s)}(p^2) &=& \frac{\sqrt{\lambda (p, m_A, m_B)}}{64 \pi}
\lambda_V^i \lambda_V^j \left[\frac{(p^2-m_A^2 - m_B^2)^2}{m_A^2
m_B^2} +8\right]\,. \eeq

In a similar way for the vector amplitudes we have: \beq
\Sigma_f^{(V)} (p^2)&=& \frac{\sqrt{\lambda (p, m_1, m_2)}}{16 \pi}
\left[ \frac{\lambda_L^i \lambda_L^j + \lambda_R^i \lambda_R^j}{3}
\left( 2 p^2 -
m_A^2 - m_B^2 - \frac{(m_A^2 - m_B^2)^2}{p^2} \right) + \right. \nonumber \\
& & \left. + (\lambda_L^i \lambda_R^j + \lambda_R^i \lambda_L^j) 2 m_A m_B \right]\,,\\
\Sigma_s^{(V)}(p^2) &=& \frac{\sqrt{\lambda (p, m_A, m_B)}}{16 \pi}
\frac{\lambda_s^i \lambda_s^j}{3} \left[ p^2 - 2 (m_A^2 + m_B^2 ) +
\frac{(m_A^2 - m_B^2 )^2}{p^2} \right]\,,\\
\Sigma_V^{(V)}(p^2) &=& \frac{\left(\lambda (p, m_A, m_B)\right)^{3/2}}{192 \pi} \lambda_V^i \lambda_V^j  \frac{p^6}{m_A^2 m_B^2} \nonumber\\
 & & \cdot \left[1 + 10 \frac{m_A^2 + m_B^2}{p^2} + \frac{m_A^4 + m_B^4 + 10 m_A^2 m_B^2}{p^4} \right]\,.
\eeq

\end{document}